\documentclass[doublecol]{epl2} 

\everymath{\displaystyle}
\usepackage{graphicx}

\usepackage{subfig}

\usepackage{amsmath,amssymb,mathrsfs}

\newcommand{\bea}{\begin{eqnarray}}
\newcommand{\eea}{\end{eqnarray}}

\title{On a G\"odel-like Solution in Non-Relativistic Gravity}
\shorttitle{On a G\"odel-like Solution in Non-Relativistic Gravity} 

\author{A. F. Santos\inst{1}\thanks{E-mail: \email{alesandroferreira@fisica.ufmt.br}}, R.G.G. Amorim\inst{2,3,4} \thanks{E-mail: \email{ ronniamorim@gmail.com}}, K.V.S. Ara\'ujo\inst{5} \thanks{E-mail: \email{ kayo.vaz@aluno.unb.br}} \and S. C. Ulhoa\inst{4,5} \thanks{E-mail: \email{ sc.ulhoa@gmail.com}} }

\shortauthor{A. F. Santos, R.G.G. Amorim, K.V.S. Ara\'ujo  and S. C. Ulhoa}

\institute{                    
  \inst{1} Programa de P\'{o}s-Gradua\c{c}\~{a}o em F\'{\i}sica, Instituto de F\'{\i}sica, Universidade Federal de Mato Grosso, Cuiab\'{a}, Brasil.\\
  \inst{2}Gama Faculty, University of Brasília,
72.444-240, Brasília, DF, Brazil.\\
  \inst{3}International Center of Physics, Instituto de F\'isica, Universidade de Bras\'ilia, 70910-900, Bras\'ilia, DF, Brazil.\\
  \inst{4} Canadian Quantum Research Center, 
204-3002 32 Ave Vernon, BC V1T 2L7,  Canada.\\
\inst{5}Instituto de F\'isica, Universidade de Bras\'ilia, 70910-900, Bras\'ilia, DF, Brazil.
}


\abstract{
The article deals with G\"odel-like solutions in the context of Galilean gravity,
a geometric formulation of non-relativistic gravitation defined on a
five-dimensional Galilean manifold.
Within this framework, non-relativistic matter fields admit a covariant
description, while the physical Newtonian dynamics is recovered through an
immersion into the usual $3+1$ spacetime.
By adopting a G\"odel-like metric ansatz and coupling the gravitational field to
a Galilean fluid derived from a variational principle, we obtain a system of
highly nonlinear and coupled field equations.
Exact solutions are constructed by fixing the matter sector consistently with
the field equations. The resulting configurations describe rotating non-relativistic universes and
satisfy $D(x)>H(x)$ throughout the entire spatial domain.
As a consequence, the associated Killing vector remains spacelike everywhere
and no closed timelike curves arise.
}

\begin{document}

\maketitle

\section{Introduction} \label{sec.1}

Causality has been a cornerstone of scientific reasoning since the very foundations of natural philosophy. It underlies the consistency of physical hypotheses and has never been empirically violated, remaining an essential guiding principle of scientific practice. With the advent of general relativity, however, this status became less secure. The existence of solutions containing singularities, such as black holes, forces one to confront causality not by discarding it, but by restricting attention to regions where causal order is preserved. In this context, G\"odel’s 1949 solution \cite{Godel1949} occupies a peculiar position: although fully consistent within Einstein’s equations, it predicts the existence of closed timelike curves (CTCs), along which the very notion of cause and effect ceases to be well defined.

Within general relativity, the strategy of avoiding pathological regions rather than discarding entire solutions has become not only acceptable but effectively standard. The presence of event horizons, for instance, limits the causal influence of singular regions without invalidating the underlying spacetime. Even G\"odel’s solution, despite its prediction of closed timelike curves, has been extensively reexamined with the aim of identifying causally well-behaved regions, as explored in works such as those by Rebouças and Tiomno~\cite{reboucas_tiomno_1983}. In this way, causality violation in G\"odel-type universes has evolved from an exotic curiosity into a serious subject of investigation, opening the door to the systematic study of concepts traditionally confined to science fiction, such as time travel. The situation in Newtonian physics, however, is fundamentally different. There, the structure of the theory is built upon an absolute notion of time, leaving no conceptual room for the breakdown of causal ordering in physical events.

The apparent rigidity of Newtonian physics with respect to causality, however, stems less from an intrinsic limitation on the propagation of information than from the structure of Galilean transformations themselves, in contrast to the relativistic case where causality is directly tied to a finite invariant speed. This observation naturally raises the question of whether a covariant structure exists for Galilean transformations. Such a question has led to several covariant formulations of non-relativistic physics \cite{LevyLeblond1967, LevyLeblond1973, takahashi_2, takahashi_3, Santana2000, Santana2003a, Santana2003b, Santana2003c}. Galilean covariance consists precisely in a covariant description of non-relativistic fields, which requires these fields to be defined on a five-dimensional Galilean manifold. The usual Newtonian $3+1$ description is recovered through an embedding specified by the choice $s=\bar{r}^{2}/(2t)$, where $s$ denotes the fifth coordinate of the Galilean manifold. Within this framework, a non-relativistic gravitational theory in close analogy with general relativity can be constructed by attributing curvature to this Galilean manifold \cite{ulhoa2009}. It is worth noting that not all approaches to a metric formulation of Newtonian gravity rely on Galilean manifolds, as exemplified by the formulation of Duval et all \cite{duval1985}; nevertheless, the Newtonian gravitational dynamics must be consistently recovered in any viable theory. It is precisely in this context that a G\"odel-type solution of Galilean gravity becomes relevant. In the following, we investigate such a solution and explore its implications for causality violation within a non-relativistic gravitational framework. This work focuses on the existence and structure of solutions, rather than on their dynamical or observational consequences.

This article is organized as follows. In Sec.II, we present a brief description of Galilean covariance. In Sec.III, the field equations of Galilean gravity are introduced and a G\"{o}del-type solution is obtained. The causal structure of the resulting spacetime is then analyzed using a numerical approach. Finally, our conclusions are presented in the last section. Throughout this work, natural units are adopted.

\section{Galilean Gravity} \label{sec.2}

In this section we present the covariant formulation of Galilean transformations, which in their traditional form appear as a 3+1 decomposition,
\begin{equation}
\mathbf{x}’ = R\,\mathbf{x} + \mathbf{v}\,t + \mathbf{a},
\qquad
t’ = t + \tau .
\end{equation}
In this representation, space and time are treated as quantities of distinct nature, a feature that characterizes non-relativistic physics. However, it is possible to adopt a system of natural units in which the velocity becomes dimensionless, v=1, allowing spatial and temporal coordinates to be handled on the same formal footing. This choice does not imply any relativistic interpretation of time, but rather provides the dimensional compatibility required for a covariant description of coordinate transformations of different physical character.

Within this framework, non-relativistic transformations can be written in a covariant form by introducing a five-component coordinate vector,
\begin{equation}
x^\mu = (x^1,x^2,x^3,x^4,x^5)
\equiv (\mathbf{x},t,s).
\end{equation}
These coordinates belong to a Galilean manifold and admit an inner product defined as
\begin{equation}
(x|y)=x^\mu y_\mu
= \mathbf{x}\cdot\mathbf{y} - x^4 y^5 - x^5 y^4 .
\end{equation}
This inner product defines the metric tensor
\begin{equation}
g_{\mu\nu}=
\begin{pmatrix}
\delta_{ij} & 0 & 0 \\
0 & 0 & -1 \\
0 & -1 & 0
\end{pmatrix}\,,
\end{equation}
such that
\begin{equation}
(x|y)=g_{\mu\nu}x^\mu y^\nu .
\end{equation}
In this way, a five-dimensional geometric structure, known as the Galilean manifold, is established, providing the basis for a covariant formulation of non-relativistic theories.

It is instructive to understand how the traditional 3+1 Galilean transformations are recovered from the five-dimensional non-relativistic structure. For this purpose, the coordinate s is chosen according to
\begin{equation}
x^5 = s = \frac{\mathbf{x}^2}{2t}.
\end{equation}
This procedure, known as dimensional reduction by immersion, allows the covariant non-relativistic formulation to be consistently compared with the experimentally established Newtonian description. In this way, the additional coordinate does not represent an independent physical degree of freedom, but rather a geometric device required to implement the property of covariance.

It is worth noting that the coordinate transformation
\begin{equation}
x^4 = \frac{t-s}{\sqrt{2}},
\qquad
x^5 = \frac{t+s}{\sqrt{2}},
\end{equation}
brings the Galilean metric into the diagonal form $\mathrm{diag}(1,1,1,-1,1)$. In this representation, the geometric structure becomes closer to that of a five-dimensional de Sitter-like space. Written in this form, the Galilean metric becomes more familiar from the perspective of quantum field theory techniques, a feature that will be particularly useful in the interpretation of the momentum invariant discussed below.

The momentum structure on the Galilean manifold is represented by the five-vector
\begin{equation}
p^\mu = (-i\nabla,-i\partial_t,-i\partial_s)
= (\mathbf{p},-E,-m).
\end{equation}
It is important to note that the invariant is
$p^\mu p_\mu = \mathbf{p}^2 - 2 E m$.
Therefore, imposing the condition $p^\mu p_\mu = 0$ immediately yields
$E = \frac{\mathbf{p}^2}{2m}$,
which is precisely the dispersion relation of a free particle in classical (non-relativistic) physics. In this sense, a null invariant in momentum space encodes the non-relativistic energy relation.

From the five-dimensional perspective, this condition admits a geometric interpretation: the particle is constrained to lie on the light cone of the effective de Sitter-like space associated with the Galilean manifold. This interpretation will play an important role in the formulation of non-relativistic field dynamics in a Galilean background. To this end, we consider a Klein-Gordon field defined on this manifold, whose dynamics is governed by
\begin{equation}
\partial_\mu \partial^\mu \Psi = 0\,,
\end{equation}
which corresponds to a massless scalar field in five dimensions. It is important to emphasize that the momentum canonically conjugate to the coordinate s is not the mass of the scalar field itself. Instead, this component satisfies the eigenvalue equation
\begin{equation}
\partial_s \Psi = -\, i m\, \Psi .
\end{equation}
Therefore, by adopting the separation of variables
\begin{equation}
\Psi(\mathbf{x},t,s)=e^{-i m s}\,\psi(\mathbf{x},t)\,,
\end{equation}
one obtains
\begin{equation}
i\,\partial_t \psi
-\frac{1}{2m}\,\nabla^2 \psi =0\,,
\end{equation}
which is precisely the Schr\"odinger equation for a free particle of mass m. In other words, the massless Klein-Gordon field in five dimensions provides an exact representation of the Schr\"odinger equation without interaction.

More generally, this result illustrates a central aspect of the Galilean covariant framework, non-relativistic fields can be consistently described in terms of relativistic-like fields defined on a five-dimensional manifold. In this correspondence, the covariant structure of the Galilean metric (which can be brought to a de Sitter-like form by a suitable choice of coordinates) allows non-relativistic dynamics to be formulated using the same geometric and field-theoretical tools employed in relativistic theories. This perspective will be crucial in the construction of interacting non-relativistic systems, such as fluids described by the Navier-Stokes equation, within the Galilean covariant approach.

Consider now the following Lagrangian density defined on the Galilean manifold,
\begin{equation}
\tilde{\mathcal{L}}[\rho,\phi]
=
-\frac{1}{2}\,\rho\,\partial_\mu\phi\,\partial^\mu\phi
- V(\rho)\,,
\label{lag}
\end{equation}
which leads to the field equation
\begin{equation}
\frac{1}{2}\,\partial_\mu\phi\,\partial^\mu\phi
+ V'(\rho)=0\,.
\label{navier5d}
\end{equation}
After performing the dimensional reduction associated with the Galilean immersion, this equation can be rewritten as
\begin{equation}
\partial_t \phi
+ \frac{1}{2}\,\nabla\phi\cdot\nabla\phi
= -\,V'(\rho)\,.
\end{equation}
By identifying the velocity field as $\mathbf{v}=\nabla\phi$ and introducing the continuity equation
\begin{equation}
\partial_t \rho
+ \nabla\!\cdot(\rho\,\mathbf{v})=0\,,
\end{equation}
one finds that the field equation (\ref{navier5d}) becomes
\begin{equation}
\partial_t \mathbf{v}
+(\mathbf{v}\!\cdot\!\nabla)\mathbf{v}
=
-\frac{1}{\rho}\,\nabla p(\rho)\,.
\end{equation}
This is precisely the Navier--Stokes equation in its usual form, once the immersion leading to the physical \(3+1\) formulation is adopted~\cite{Santana2003b}. Therefore, a Galilean fluid can be consistently described by the Lagrangian density (\ref{lag}). The corresponding energy--momentum tensor is given by
\begin{equation}
T^{\mu}{}_{\nu}
=
\frac{\partial \mathcal{L}}
{\partial(\partial_\mu\phi)}
\,\partial_\nu\phi
-\delta^\mu{}_\nu\,\mathcal{L}\,,
\end{equation}
which explicitly reads

\begin{equation}
\begin{split}
T_{\mu\nu}
&= \rho \Biggl[
\frac{1}{2} (\partial_\alpha \phi) (\partial^\alpha \phi) \, g_{\mu\nu}
- (\partial_\mu \phi) (\partial_\nu \phi)
\Biggr] \\
&\quad - V(\rho) \, g_{\mu\nu}.
\label{fluidogal}
\end{split}
\end{equation}
This tensor satisfies identically the conservation law
\begin{equation}
\partial_\mu T^{\mu}{}_{\nu}=0\,.
\end{equation}
Hence, a Galilean fluid is described in five dimensions by a scalar field with an interaction potential. In what follows, we adopt the separation of variables

\begin{equation}
\begin{split}
\phi(x,y,z,t,s) &= \varphi(t) + \psi(x,y,z), \\
\rho(x,y,z,t,s) &= \Theta(t,s) \, \sigma(x,y,z),
\end{split}
\end{equation}
which proves to be convenient for the coupling between Galilean gravity and a non-relativistic fluid.

It is worth emphasizing that, in its standard Newtonian formulation, the Navier--Stokes equation is not naturally obtained from a variational principle. In contrast, within the Galilean covariant framework, the fluid dynamics follows directly from a scalar Lagrangian density defined on the five--dimensional manifold. This illustrates how Galilean covariance provides a geometric reinterpretation of classical physics, allowing non--relativistic dynamics to be reformulated in a way closely analogous to relativistic field theories.

\section{G\"odel-type Solution in Galilean Gravity} \label{sec.3}

In this section we introduce Galilean gravity as a geometric formulation of non--relativistic gravitation. In this approach, Newtonian gravity is reformulated as a five--dimensional covariant theory defined on the Galilean manifold, in close analogy with the geometric description of gravity provided by General Relativity. The guiding principle is the same: gravitation is encoded in the curvature of spacetime. The crucial difference, however, is that the flat background is not the Minkowski spacetime, but the Galilean manifold itself, which naturally accommodates covariant Galilean transformations instead of Lorentz transformations. 

As discussed in the previous section, non--relativistic physical systems can be described covariantly by relativistic--like field theories defined on the Galilean manifold. This principle extends consistently to the gravitational interaction. Galilean gravity therefore provides a geometric description of Newtonian gravitation in which the dynamical degrees of freedom of the gravitational field are associated with the curvature of a five--dimensional Galilean spacetime. The corresponding field equations are obtained by equating geometric tensors constructed from the Galilean metric to the energy--momentum tensor of the non--relativistic matter fields, derived from the appropriate Galilean covariant Lagrangian density~\cite{ulhoa2009}. The field equations therefore assume the same formal structure as the Einstein equations written in five dimensions, namely
\begin{equation}
G_{\mu\nu} = 8\pi\, T_{\mu\nu},
\end{equation}
where \(G_{\mu\nu}\) is the Galilean Einstein tensor and natural units are adopted. It is important to recall that, once the immersion to the physical \(3+1\) dynamics is performed, the standard Newtonian theory of gravitation is consistently recovered.

Our goal is not to analyze particle motion, but to establish the existence of Gödel-like solutions in Galilean gravity. Thus the line element of a five-dimensional G\"odel-type space can be written as~\cite{Reb99}
\begin{equation}
dS^2=
\bigl[dt+H(x)\,dy\bigr]^2
- dx^2
- D^2(x)\,dy^2
- F^2(s)\,dz^2
- ds^2 \,,
\label{metric_godelgal}
\end{equation}
where the five-dimensional invariant interval is denoted by $S$, and the coordinate introduced to implement non-relativistic covariance is denoted by $s$. From this line element, the field equations are computed and listed below.

\begin{align}
\label{Eq:4}
&4 D(x) \left[ 4\pi \Theta(s) \sigma(x) \bigl( p(x)^{2} + 2 \bigr) - \frac{\ddot{F}(s)}{F(s)} \right] \notag\\
&+ \frac{3 H'(x)^{2}}{D(x)} = 4 D''(x), \\
\label{Eq:5}
&2 D(x) \Bigl[ - \frac{2 H(x) \ddot{F}(s)}{F(s)} + H''(x) \notag\\
&+ 8\pi H(x) \Theta(s) \sigma(x) \bigl( p(x)^{2} + 2 \bigr) \Bigl] + \frac{3 H(x) H'(x)^{2}}{D(x)} \notag \\
&\quad = 4 H(x) D''(x) + 2 D'(x) H'(x), \\
\label{Eq:6}
&\frac{H'(x)^{2}}{4 D(x)^{2}} + \frac{\ddot{F}(s)}{F(s)} = -4\pi \Theta(s) \sigma(x) \bigl( p(x)^{2} - 2 \bigr), \\
\label{Eq:7}
&\frac{D(x)^{2} \ddot{F}(s)}{F(s)} + \left( \frac{3 H(x)^{2}}{4 D(x)^{2}} + \frac{1}{4} \right) H'(x)^{2} + H(x) H''(x) \notag \\
&\quad = \frac{H(x) \bigl[ H(x) D''(x) + D'(x) H'(x) \bigr]}{D(x)} \notag\\
&+ 4\pi \Theta(s) \sigma(x) \bigl( D(x)^{2} - H(x)^{2} \bigr) \bigl( p(x)^{2} + 2 \bigr) + \frac{H(x)^{2} \ddot{F}(s)}{F(s)}, \\
\label{Eq:8}
&\frac{F(s)}{D(x)} \Bigl[ -4 D(x) D''(x) + 16\pi D(x)^{2} \Theta(s) \sigma(x) \bigl( p(x)^{2} + 2 \bigr)\notag\\
&+ H'(x)^{2} \Bigr] = 0, \\
\label{Eq:9}
&\frac{H'(x)^{2}}{D(x)} + 16\pi D(x) \Theta(s) \sigma(x) \bigl( p(x)^{2} + 2 \bigr) = 4 D''(x)\,,
\end{align}
where the energy--momentum tensor employed is that of the Galilean fluid given in Eq.~(\ref{fluidogal}). The notation adopted is as follows: a prime denotes differentiation with respect to the spatial coordinate $x$, while a dot denotes differentiation with respect to the Galilean coordinate $s$. Moreover, the pressure $p$ is defined as the gradient of the scalar field $\psi$. In this work, the term ``cosmological'' is used in the same sense as in G\"odel’s original solution, namely as a homogeneous rotating spacetime, without invoking expansion or isotropy.

It is important to stress that the field equations (\ref{Eq:4})--(\ref{Eq:9}) were obtained by adopting the linear potential
\begin{equation}
V(\rho)=\rho ,
\end{equation}
in the Galilean fluid Lagrangian. This choice implies $V'(\rho)=\mathrm{const.}$ and, consequently, a vanishing pressure gradient in the reduced $3+1$ dynamics, corresponding to a fluid with constant pressure, since
$\nabla p = \rho \nabla V'(\rho)=0$. Due to the redundancy of the field equations, one of the functions may be fixed
without loss of generality. We therefore choose $\Theta=\text{const}$ as a
convenient normalization. This choice is consistent with the structure
of the Galilean fluid employed and does not restrict the remaining dynamical
degrees of freedom.

In the numerical analysis that follows, we further fixed $\Theta(s)=1$ and $p(x)=1$. Although these choices were initially implemented for simplicity, they are in fact consistent with the continuity equation and with the Bernoulli--type equation derived from the Galilean fluid dynamics. In particular, once $V(\rho)=\rho$ is assumed, the pressure must be constant, so that the adopted configuration represents the only admissible solution within this class. Therefore, the subsequent solutions do not rely on arbitrary gauge fixing, but rather reflect a consistent realization of the Galilean fluid coupled to gravity.

\begin{figure}[htbp]
\centering
\subfloat[$F(s)$ and $\Theta(s)$.\label{fig:FTheta}]{
\includegraphics[width=0.48\textwidth]{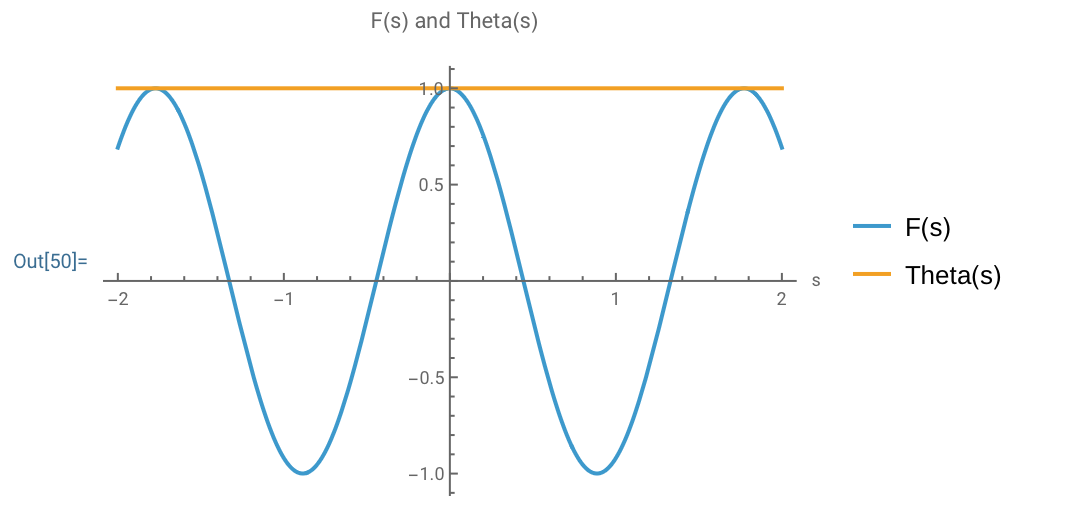}
}\hfill
\subfloat[$D(x)$ and $H(x)$.\label{fig:HD}]{
\includegraphics[width=0.48\textwidth]{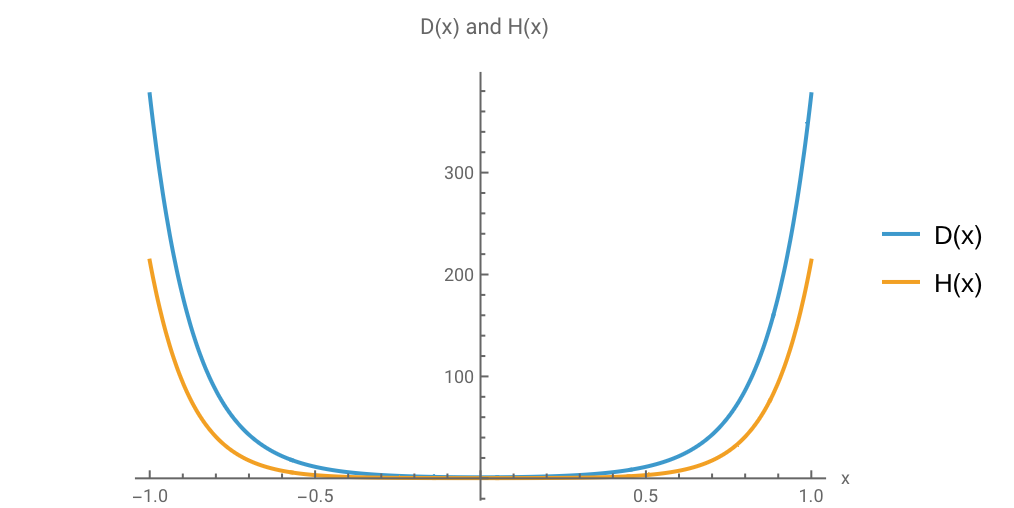}
}\\[0.8em]
\subfloat[$\sigma(x)$.\label{fig:sigma}]{
\includegraphics[width=0.48\textwidth]{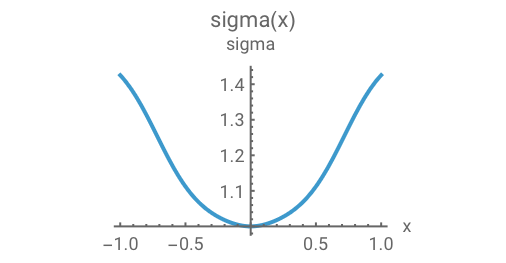}
}\hfill
\subfloat[$p(x)$.\label{fig:px}]{
\includegraphics[width=0.48\textwidth]{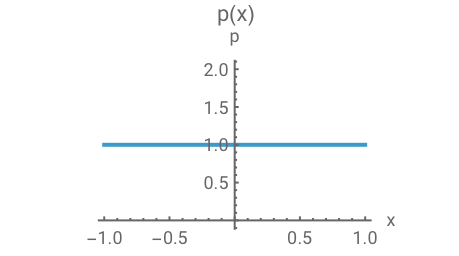}
}
\caption{Numerical profiles obtained for the G\"odel-like Galilean ansatz.}
\label{fig:solutions}
\end{figure}

Figures~\ref{fig:solutions} summarize the numerical profiles associated with the G\"odel-like ansatz in Galilean gravity. 
In Fig.~\ref{fig:HD} the functions $D(x)$ and $H(x)$ satisfy $D(x)>H(x)$ throughout the plotted domain. 
Since the metric component along the $y$-direction is $g_{yy}=H(x)^{2}-D(x)^{2}$, this implies $g_{yy}<0$ everywhere in the same domain, so the Killing vector $\partial_{y}$ remains spacelike and no noncausal region emerges from the ansatz at this level. The function $F(s)$ shown in Fig.~\ref{fig:FTheta} follows the same qualitative class of solutions for the extra coordinate dependence reported in Ref.~\cite{Reb99}, as expected from the structure of the field equations where the $s$-sector is constrained by a second-order relation involving $\ddot F/F$. 
In the present solution, $\Theta(s)$ is taken constant, which simplifies the matter sector and isolates the behavior of $F(s)$ associated with the extra coordinate. The spatial density profile encoded in $\sigma(x)$ is displayed in Fig.~\ref{fig:sigma}. With $\Theta$ fixed, $\sigma(x)$ directly sets the $x$-dependence of the matter distribution entering the field equations, and the resulting profile is smooth and symmetric in the plotted range. 
Finally, Fig.~\ref{fig:px} shows that the pressure is constant in this configuration, consistent with the simplifying assumption adopted for the numerical construction and useful as a baseline for exploring more general fluid profiles.

\section{Conclusion} \label{sec.4}

In this article we proposed a G\"odel-like solution within the framework of
Galilean gravity.
The construction follows the standard form discussed in the literature for
relativistic G\"odel-type spacetimes, now applied to a genuinely
non-relativistic gravitational theory formulated in five dimensions.
The Galilean gravitational field equations were consistently coupled to the
energy--momentum tensor of a Galilean fluid, which reproduces the
Navier--Stokes equation under immersion. The resulting field equations reduce to a highly nonlinear coupled system for
six independent functions.
By fixing the potential as $V(\rho)=\rho$ and imposing constant $\Theta$ and
pressure, we obtained a class of solutions compatible with the Galilean fluid
dynamics and with the assumed symmetry of the metric.
The spatial profiles of the metric functions $H(x)$ and $D(x)$ play a central
role in the causal structure of the model.

The sector associated with the extra Galilean coordinate $s$ decouples from the
causal analysis.
The function $F(s)$ satisfies the same class of equations found in previous
studies of G\"odel-type solutions, while the physically relevant Newtonian
density is entirely determined by the spatial function $\sigma(x)$ in the
$3+1$ description.
This reinforces the interpretation of Galilean gravity as a geometric theory
capable of reproducing non-relativistic physics without introducing causal
pathologies. 

A central result of this work is that the solutions obtained satisfy
$D(x) > H(x)$ over the entire spatial domain.
As a direct consequence, the Killing vector $\partial_y$ remains spacelike
everywhere and no closed timelike curves arise in this Galilean G\"odel-like
spacetime.
In contrast to its relativistic counterpart, in which causality violation is an
intrinsic feature of the geometry, the present non-relativistic model describes
a fully causal rotating universe.
This result indicates that the breakdown of causality in G\"odel-type geometries
is not a mere consequence of covariance itself, but is instead deeply connected
to the relativistic causal structure of spacetime.

The absence of closed timelike curves in the present solution therefore carries
a broader conceptual significance.
In relativistic gravity, G\"odel-type spacetimes are commonly regarded as
pathological and are consequently excluded from cosmological modeling due to the
presence of closed timelike curves.
The Galilean counterpart constructed here demonstrates that this obstruction is
not intrinsic to rotating cosmological configurations as such, but rather a
direct consequence of the relativistic spacetime structure.
Within a genuinely non-relativistic gravitational framework, G\"odel-like
solutions emerge as fully causal and thus physically admissible configurations.
This places rotating Galilean universes on an equal footing with standard
cosmological models as potential descriptions of anisotropic phenomena.
From this perspective, Galilean G\"odel-like spacetimes may provide an
alternative geometric setting for the analysis of cosmological data exhibiting
large-scale anisotropies, without encountering the conceptual difficulties
associated with causality violation.
These results suggest that the systematic exclusion of G\"odel-type geometries
from cosmological considerations should be reconsidered in the non-relativistic
regime. Future investigations may explore the dynamical behavior of test particles and
fluid perturbations in such backgrounds, as well as possible observational
signatures and the role of more general matter contents and symmetry assumptions
within the Galilean gravitational framework.

\acknowledgments

This work by A. F. S. is partially supported by National Council for Scientific and Technological
Development - CNPq project No. 312406/2023-1.



\end{document}